\documentclass[preprint, authoryear,11pt]{elsarticle}
\usepackage{amssymb}
\usepackage{color}
\usepackage{amsmath}
\usepackage{subfigure}
\usepackage{graphicx}
\usepackage{graphics}
\usepackage{booktabs}
\usepackage{threeparttable}
\usepackage{appendix}
\usepackage{multirow}
\usepackage{geometry}
\usepackage{setspace}
\usepackage[colorlinks,linkcolor=red,anchorcolor=blue,citecolor=green]{hyperref}
\usepackage{soul}
\usepackage{subfigure} 
\usepackage{mathpazo}
\usepackage{algorithm}  
\usepackage{algorithmicx}  
\usepackage{algpseudocode}  
\usepackage{xcolor}
\usepackage{booktabs}
\usepackage{hyperref, tabularx}

\usepackage{amsmath}

\newif\ifshowshreyaa
\showshreyaafalse 

\ifshowshreyaa
  \newcommand{\shreyaa}[1]{\textcolor{blue}{#1}}
\else
  \newcommand{\shreyaa}[1]{}
\fi

\journal{Transportation Research Part C}

\begin{document}
\begin{spacing}{1.3}

\begin{frontmatter}

\title{Making Highway Performance Measures Actionable: A Controllability-Aware Measure of Traffic Congestion}


\author[mit]{Shreyaa Raghavan\corref{cor1}}
\ead{shreyaar@mit.edu}
\author[mit]{Edgar Ramirez-Sanchez}
\author[mit]{Monica Chan}
\author[mit]{Zhengbing He\fnref{present}}
\author[mit]{Cathy Wu}

\address[mit]{Laboratory for Information \& Decision Systems (LIDS), Massachusetts Institute of Technology, Cambridge, MA 02139, USA}

\cortext[cor1]{Corresponding author}
\fntext[present]{Present address: Department of Electrical and Electronic Engineering,
                 University of Nottingham Ningbo China, 199 Taikang East Road,
                 Ningbo 315104, Zhejiang, China}
                 
\begin{abstract}
Emerging intelligent transportation systems (ITS) show substantial benefit in reducing congestion, emissions, and accidents and enable lower-cost alternatives to highway lane expansion. However, there currently exists no actionable metric by which agencies can assess the improvement potential from these interventions and then invest in infrastructure where it will be most promising. To guide ITS deployment, we introduce \emph{controllability-aware performance measures} that quantify the upper bound of improvement that speed-based vehicle control can achieve under fixed demand. We operationalize this with \emph{controllable congestion}, the maximum achievable reduction in system delay from controlling speed limits, which we estimate via a locally optimal receding-horizon optimization framework. This nonlinear optimization framework is grounded in a smooth reformulation of the METANET macroscopic traffic model and solved using model predictive control (MPC). Using both a synthetic scenario and highway data from the I-24 SMART Corridor in Tennessee, we find that controllable congestion is largely independent of total delay, revealing that for two days with nearly identical travel times, controllable congestion varies from 35\% to 86\%. We further show how the realizable share of this upper bound depends on the operational constraints imposed. On I-24, the minimum posted speed limit has little effect on controllable congestion, while the frequency of speed limit updates and the maximum difference between adjacent gantries have large impacts. This framework allows highway operators to distinguish between congestion that is structurally constrained by capacity and congestion that is highly responsive to ITS, enabling more cost-effective deployment of control-based traffic infrastructure.

\end{abstract}

\begin{keyword}
Controllable congestion \sep Freeway traffic control \sep Performance measures \sep Variable speed limits \sep Model predictive control \sep METANET

\end{keyword}

\end{frontmatter}

\newpage




\newpage

\section{Introduction}

\shreyaa{
[Performance measures are the key lever for stakeholders to make important decisions about infrastructure]} 

As in most systems, performance measures serve as a key lever in the management of transportation infrastructure, allowing stakeholders to take informed and strategic actions with respect to planning, investment, and operations. By quantifying various aspects of system performance, performance measures provide objective measures of success for stakeholders and help them assess the quality of critical transportation systems, conveying whether a system is in a ``good" (i.e. desirable) state or a ``bad" (i.e. undesirable) state. In highway systems, approaches grounded in performance measures or aimed at improving them have increasingly shaped policy, funding decisions, and infrastructure development. As such, the selection and usage of appropriate performance measures is a significant part of maintaining and improving highway systems. 

To assess system performance, most existing metrics use historical data to describe the current or past state of the system. For instance, traffic congestion is typically evaluated using measures such as total delay or peak-hour excessive delay (PHED) to indicate how congested a corridor is \citep{margiotta2018national}. Similarly, highway safety is often assessed through crash statistics, including the number of fatalities and injuries \citep{FHWA_StateSafetyPerformanceTargets_2025}. By benchmarking against perfect conditions (e.g., zero delay or zero crashes), these metrics quantify how far a system is from the absolute ideal. While useful for diagnosing the performance of the system, such metrics do not convey how much the system can be improved in actuality. Since they are derived from historical data, they are fundamentally retrospective and used to infer performance rather than to assess the system’s potential responsiveness to control or management strategies. In other words, they measure \emph{performance outcomes}, but not \emph{performance potential}. As Figure 1 illustrates, measuring traffic congestion, on its own, does not tell an agency what highway infrastructure to invest in. 

\begin{figure}[H]
    \centering
    \includegraphics[width=0.9\linewidth]{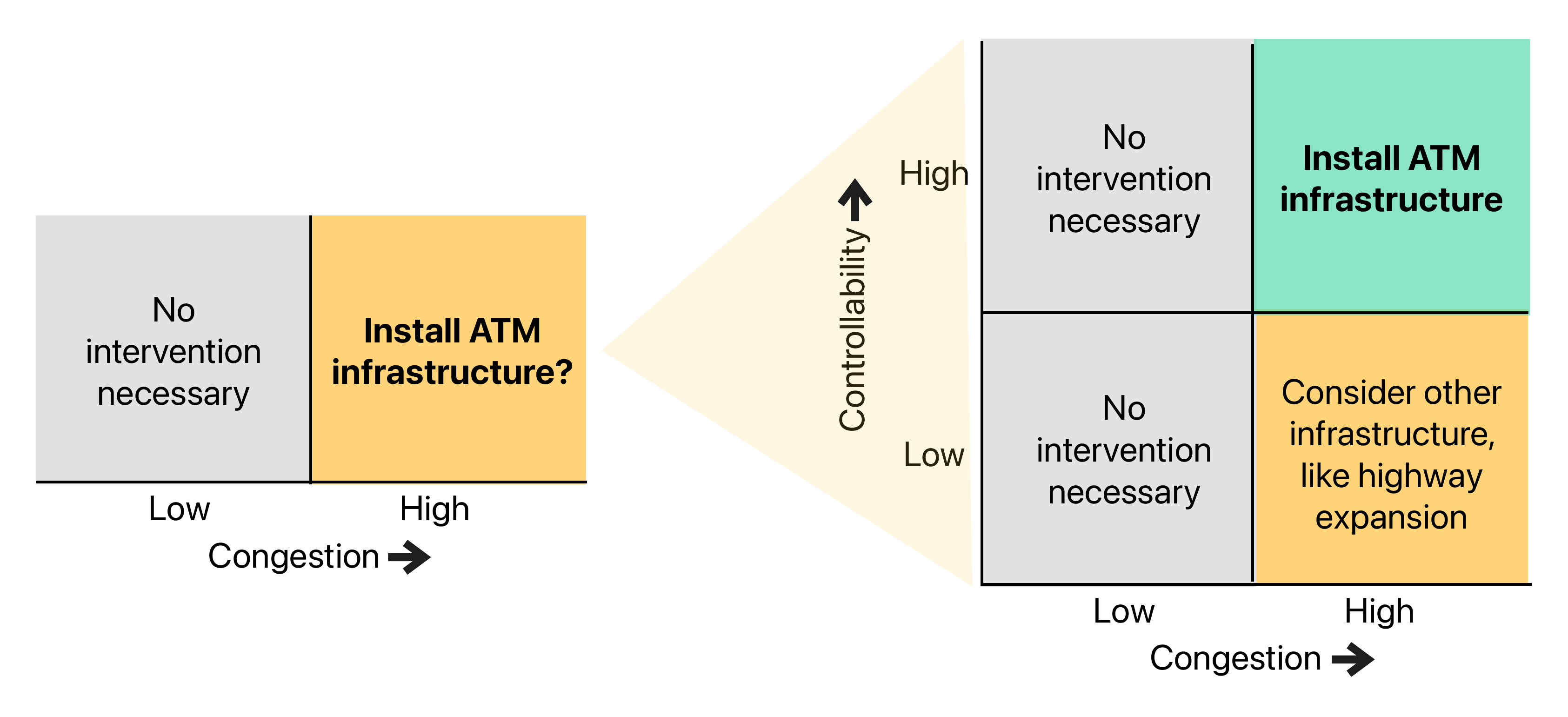}
    \caption{A decision framework showing how existing congestion metrics combined with controllability can determine whether no action is needed, ATM infrastructure should be installed, or highway expansion should be considered.}
    \label{fig:con-venn}
\end{figure}

The ability to quantify performance potential is increasingly important as intelligent transportation systems (ITS) --- such as autonomous vehicles (AVs), variable speed limits (VSL), and adaptive cruise control (ACC) --- offer low-cost alternatives to traditional congestion mitigation strategies like highway expansion. These emerging interventions, also known as active traffic management (ATM) strategies, act as control actuators to the system, aiming to dynamically coordinate and optimize the flow of \emph{existing} traffic based on the current traffic conditions \citep{mirshahi2007active}. These technologies have shown success in improving many highway systems through field experiments \citep{lee2025traffic, kreidieh2018dissipating, wang2016connected}. Notably, they are also often significantly less costly than traditional highway expansion projects. For example, the Michigan Department of Transportation (DOT) reports that two VSL gantries spaced out over one mile cost \$1 million, while constructing a single lane of an urban freeway for the same length is estimated at \$7.7 million \citep{jenior2019decision, strongtowns2020mile}. However, existing metrics provide no way to preemptively assess how much benefit these smarter, speed-control management strategies provide. 

A more meaningful benchmark would capture how \emph{controllable} a system is, which we define as the extent to which a system’s state can be moved toward a better-performing regime. Unlike its formal definition in control theory, here, we refer to controllability in an operational sense. In the context of highway management, this translates to how much a highway’s performance could theoretically improve under optimal ATM strategies. Without this perspective, decision-makers may lack the ability to distinguish between highways that are simply performing poorly and those that are both poor and \emph{improvable}. For instance, two equally congested freeways may have different potential for improvement depending on various features such as demand and driving behavior. As shown in Figure \ref{fig:con-venn}, only when congestion and controllability are both high do we expect interventions to yield substantial benefits. In contrast, high congestion with low controllability suggests that improvements are unlikely with ATM, while high controllability with low congestion indicates limited need for intervention. Without systematically quantifying this potential controllability alongside existing metrics, agencies risk deploying resources in ineffective locations.

To address this gap, we introduce and formalize the concept of controllability for highway systems and develop a corresponding metric --- controllable congestion --- that quantifies the maximum achievable reduction in delay under idealized, speed-based traffic management. This benchmark is not intended as a real-time controller, but rather as a planning tool that reveals the upper bound of how much improvement is theoretically attainable with ATM strategies. To estimate this upper bound, i.e. controllable congestion, we propose a nonlinear optimization framework based on the METANET macroscopic model and state-of-the-art nonlinear model predictive control (NLMPC). Through simulation studies and real data from the I-24 SMART Corridor, we show that controllable congestion is highly non-monotonic with respect to demand, revealing that control is largely ineffective in extremely high demand scenarios. We also conduct a thorough analysis on how controllable congestion varies with respect to system constraints and share insights on how it can drastically change the effectiveness of deployed speed limit control. This framework allows agencies to distinguish between congestion that is \emph{structurally constrained by capacity} and congestion that is \emph{highly responsive to ATM}, ultimately enabling smarter, more cost-effective deployment of intelligent traffic management.

Below, we summarize the main contributions of this work:
\begin{itemize}
    \item We propose and formalize controllability-aware performance measures as an \emph{actionable} counterpart to existing performance metrics, capturing the previously unquantified potential of low-cost and novel ITS management strategies.
    \item We introduce controllable congestion, a standardized metric measuring the upper bound of delay that can be eliminated under perfect active traffic management. Then, we develop a model-based methodology to estimate controllable congestion, using nonlinear model predictive control (NLMPC) and the macroscopic traffic model, METANET.
    \item Through simulation studies and real data from the I-24 SMART Corridor, we show that controllable congestion is largely independent of total delay: for two days with nearly identical travel times, controllable congestion varies from 35\% to 86\%.
    \item We conduct a thorough analysis on how controllable congestion varies with respect to system constraints and share insights on how specific constraints, like frequency of speed limit update, can drastically change the responsiveness of speed-control strategies. 
\end{itemize}

\section{Related Work} \label{lit_review}

\subsection{Performance measures}

To understand which projects to allocate investments toward, national and state-level agencies have long used performance measures as a way of performance-based assessment. For example, metrics for traffic congestion are primarily point measures or delay-based measures. Point measures, like level-of-service (LOS)  or volume-to-capacity (V/C) ratio, typically measure congestion via densities and velocities from loop detector or radar data \citep{dowling2007traffic}. These have a clear threshold at which operators consider the highway to be congested, e.g., LOS is categorized from A to F depending on the density or average speed \citep{los_original}. Delay-based measures, on the other hand, capture how much congestion affects drivers' speeds or travel times and are commonly derived from probe data. Examples of this are average delay,  travel time index (TTI), and peak-hour excessive delay (PHED), which is the number of veh-hrs during which the average speed is less than 20 mph \citep{margiotta2018national}. Delay-based measures are used by major national reports of highway efficiency and performance, such as the Texas Urban Mobility Report \citep{schrank20032003}.

For safety, national practice is guided by data on fatalities, serious injuries, and accident rates per VMT \citep{national2004fatality}. Surrogate safety measures, which are observable crash-related metrics, provide an additional suite of metrics that allow for simulation-based safety analysis. Examples of this are Time-to-Collision (TTC) and Post-Encroachment Time (PET), which anticipate potential safety risks before collisions occur \citep{Porter_et_al_2019_SafetyAnalysisTSMO}. On the other hand, performance measures for emissions typically focus on quantifying the amount of pollutants produced by vehicles over a given time or distance. Commonly tracked pollutants include carbon dioxide (CO$_2$), nitrogen oxides (NO$_{\text{x}}$), and particulate matter (PM$_{2.5}$ and PM$_{10}$). These measures are expressed as total emissions or emissions per vehicle-mile traveled (VMT). Calculations are often performed using models such as the EPA’s Motor Vehicle Emission Simulator (MOVES), which translates traffic activity (volumes, speeds, fleet mix) into pollutant outputs \citep{vallamsundar2011overview}.

For large-scale infrastructure projects, agencies typically evaluate improvement by comparing a collection of these performance measures before and after deploying a management strategy. These measures are selected based on criteria such as significance, measurability, and data availability \citep{margiotta2024operations}. For instance, a project optimizing traffic signals along Baltimore arterials used delay, number of stops, fuel consumption, and CO$_2$ emissions as performance metrics \citep{margiotta2024operations}, while a project in Missouri and Michigan on dynamic lane merging in work zones relied on queue length, delay, and speed \citep{savolainen2024improving}. Although such measures capture important aspects of system performance, stakeholders --- including DOTs and Metropolitan Planning Organizations (MPOs) --- consistently emphasize that the most useful metrics are those that can be directly influenced through their investments and policies \citep{margiotta2024operations, fhwa_rosap_67101}. Yet many existing analyses overlook this criterion, leaving agencies uncertain, prior to deployment, on whether the targeted metrics can in fact be improved, and if so, by how much. This work addresses that gap by formally defining an axis of controllability, which provides a systematic way to preemptively estimate how much a metric could be improved. The example we consider in this work is controllable congestion, which is the upper bound of delay that can be reduced under highway speed control. The goal is for agencies to assess the extent to which their metrics of interest can be improved prior to infrastructure investment, not after.


\subsection{Traffic control infrastructure}

While performance measures describe the severity of congestion, the traffic control literature offers a complementary perspective: how management strategies can alter system dynamics. To ground our notion of controllability --- specifically controllable congestion --- we draw on work in smart and adaptive traffic control. In particular, since in the methodology we focus on management strategies related to speed limits, we look to literature on macroscopic control mechanisms where multiple vehicles (located nearby each other) are given the same control signal, in contrast to Lagrangian approaches that target individual vehicles. This type of control is typically realized and deployed as roadside variable speed limits (VSL) \citep{zhang2025real, kuhn2016evaluation} or in-vehicle signage (IVS) \citep{xu2023speed, bhattacharyya2024enhancing}. There is a substantial body of work on how to control these physical systems and what speed limits to provide vehicles, mainly from literature on VSL. These methods primarily fall into two main categories: reactive or proactive.

\emph{(1) Reactive Strategies: } Reactive methods follow heuristic or feedback-based rules --- like speed matching or threshold-based adjustments --- that respond directly to observed traffic states. These controllers are popular due to the straightforward implementation and minimal required computation. Various early studies have shown the effectiveness of even simple rule-based controllers on improving safety and reducing congestion because they enable speed harmonization \citep{smulders1990control, grumert2018characteristics, piao2008safety}. The SPECIALIST algorithm \citep{hegyi2008specialist, hegyi2010dynamic} is a notable example of feedback VSL control designed to dissipate stop-and-go waves. It employs kinematic wave theory and real-time detector data to determine appropriate speed limits during a shock wave. However, reactive strategies are limited in that there is a time delay, and by the time an action is taken, traffic may have already worsened.

\emph{(2) Proactive Strategies: } To address the short-sighted nature of reactive strategies, proactive controllers, that predict a jam and aim to preemptively control it, have gained popularity. In particular, model predictive control (MPC) has been shown to perform well \citep{hegyi2005optimal, yu2014optimal, frejo2014hybrid}. MPC uses a traffic model, typically macroscopic, to predict the effect of the current control action on future time steps and minimize the objective over a longer time horizon to avoid shortsighted actions. While most studies use METANET, a second order macroscopic traffic flow model, some have used first order models, like an extended cell transmission model \citep{han2017resolving} or a linearized version of METANET \citep{chavoshi2023feedback} to improve tractability. Model-based VSL controllers have also been optimized alongside ramp metering in various studies \citep{hegyi2002optimal, lu2010combining}. Reinforcement learning (RL)-based methods are emerging as another promising method. Although RL reacts to the current observed state, it is trained to maximize long-term cumulative reward, effectively learning anticipatory behaviors. Deep RL approaches have been shown to reduce congestion and travel time by learning policies that dynamically adjust speed limits in response to real-time traffic conditions \citep{han2022new, zhang2023marvel, zhang2025real}. For a more thorough review on both macroscopic and microscopic traffic management via speed limit control, the reader should refer to \cite{he2025review}.

Despite these advances, prior work on VSL and other macroscopic control strategies has been oriented around a fundamentally different goal than ours. Existing studies are primarily concerned with designing a specific controller and then evaluating its performance under a fixed set of assumptions, constraints, and operational rules. These studies typically report improvements such as reduced delay, improved safety, or dissipated shockwaves for a particular corridor and a particular controller configuration. As a result, their conclusions are inherently tied to the chosen control architecture and the operational constraints under which the controller is deployed.

Our work is complementary: we intentionally relax controller-specific constraints to estimate the upper bound on achievable delay reduction under any feasible speed-based management strategy. This quantity, which we define as controllable congestion, serves as a system-level property, not a controller performance statistic. In this sense, controllability acts as a precursor to controller design, providing agencies with a decision-making benchmark that identifies where management strategies are likely to be effective before deciding which controller to deploy. Methodologically, our approach shares components with model-based VSL control (e.g., MPC with METANET, as in \cite{hegyi2005optimal}), but we adopt a more general framing and emphasize this methodology as a means to quantify controllable congestion. 

\section{Preliminaries: METANET}

METANET is a popular and widely-used macroscopic traffic simulation that keeps track of the velocity, density, and flow of each highway segment at every time step \citep{messmer1990metanet}. It is a link-based, discrete-time model, meaning the highway section of interest is split into segments of equal length $L$, and each simulation time step is $\delta$ seconds. The macroscopic traffic variables --- density, velocity, and flow --- are kept track of for each segment $\ell$ and at each time step $t$ and are denoted by $\rho_t(\ell), v_t(\ell), q_t(\ell)$, respectively. All the other fixed METANET parameters are described in Table 1. These parameters may also vary by segment. The explicit dynamics equations are as follows:
\begin{align}
    \rho_{t+1}(\ell) &= \rho_{t}(\ell) + \frac{\delta}{L \lambda_\ell} \left(q_{t}(\ell-1) - \frac{q_{t}(\ell)}{1 - \beta(\ell)} + r(\ell) \right)
    \label{density} \\
    v_{t+1}(\ell) &= v_{t}(\ell) + \frac{\delta}{\tau} (V[\rho_{t}(\ell)] - v_{t}(\ell))     + \frac{\delta}{L} v_{t}(\ell) \left( v_{t}(\ell-1)- v_{t}(\ell)\right) \label{velocity} \\
    & \quad - \frac{\nu \delta}{\tau L} \frac{\rho_{t}(\ell+1) - \rho_{t}(\ell)}{\rho_{t}(\ell) + \kappa} \notag \\
     q_{t+1}(\ell) &= \rho_{t+1}(\ell) v_{t+1}(\ell) \lambda_\ell \label{flow} \\
    V[\rho_{t}(\ell)] &= \min \left( v_{\text{free}} \exp \left[-\frac{1}{\alpha}\left( \frac{\rho_{t}(\ell)}{\rho_{cr}}\right)^{\alpha}\right], u_{t}(\ell) \right) \label{V}
\end{align}
\begin{table}[!ht]
    \centering
    \caption{Parameter Definitions of METANET}
    \label{tab:parameters}
    \begin{tabularx}{0.9\textwidth}{l X}
        \toprule
        \textbf{Parameter} & \textbf{Description} \\
        \midrule
        $\delta$ & Simulation time step in seconds\\
        $L$ & Segment length in km \\
        $\tau$ & Model parameter to be calibrated \\
        $\nu$ & Model parameter to be calibrated\\
        $\kappa$ & Model parameter to be calibrated\\
        $\alpha$ & Exponential shape of fundamental diagram \\
        $v_{\text{free}}$& Free flow speed of vehicles \\
        $\rho_{cr}$& Critical density of the highway ($\rho > \rho_{cr}$ triggers congestion) \\
        $\rho_{max}$& Maximum per-lane density of the highway \\
        $Q$& Flow capacity per lane \\
        $\lambda_{\ell}$& Number of lanes in segment $\ell$\\
        $r({\ell})$& On-ramp demand for segment $\ell$; Zero if there is no on-ramp\\
        $\beta({\ell})$& Off-ramp turning ratio for segment $\ell$; Zero if there is no off-ramp\\
        \bottomrule
    \end{tabularx}
\end{table}

Equation \eqref{V}, which represents the fundamental diagram, models the effect of an enforced speed limit, denoted by $u_t(\ell)$. For the last segment, \eqref{velocity} is given a downstream boundary density for the $\rho_{t}(\ell+1)$ term. To model the upstream boundary, METANET includes a queuing term. If there is congestion at the first segment or the demand is higher than the capacity, then vehicles form a queue at the mainline origin. We introduce variables $d^{O}_t$, the mainline demand arriving at the origin at time step $t$, $q^O_t$, the true origin inflow (i.e. amount of demand that is serviced) at time step $t$, and $w_t$, the number of vehicles in the queue at time step $t$. The origin inflow replaces the upstream flow in \eqref{density} when calculating $\rho_{t+1}(1)$, so $q_t(\ell-1) = q^O_t$ when $\ell = 1$. It is the smallest of three quantities:
the demand plus queued vehicles, the capacity of the first segment, and the
flow that can enter the first segment given its current density. The equations for queuing are
\begin{align}
    q_t^O &= \min \left( d^O_t + \frac{w_t}{\delta}, ~\lambda_{1} Q, ~\lambda_{1} Q ~\frac{\rho_{max} - \rho_t(1)}{\rho_{max} - \rho_{cr}}  \right) \label{inflow}, \\
    w_{t+1} &= w_t + \delta \left(d^O_t - q_t^O\right). \label{queuing}
\end{align}

The equations \eqref{density}--\eqref{queuing} summarize the METANET dynamics. More detail on each line can be found in \cite{hegyi2005optimal}.  The simulation time step $\delta$ and segment length $L$ are set to ensure numerical stability by satisfying the Courant--Friedrichs--Lewy (CFL) condition,
\begin{equation}
    \frac{L}{\delta} \geq v_{\text{free}}.
\end{equation}
This condition ensures that the segment length is small enough such that vehicles cannot skip segments within a single time step. 

\section{Problem Formulation}

First, we consider the formal problem of quantifying \emph{controllability}, defined as the maximum achievable improvement in a system performance metric under active speed management strategies. Let $J$ denote a metric of interest (e.g., total delay, emissions, or crashes), which depends on the evolution of a highway system over some horizon $T$. We assume, without loss of generality, that a smaller value of $J$ implies better system performance, so the goal is to minimize $J.$ The system state at time $t$ is denoted by $x_t$, and evolves according to some dynamics
\begin{align}
    x_{t+1} = f(x_t, u_t, d_t) \label{metanet},
\end{align}
where $u_t$ is the control input applied at time $t$, $d_t$ represents fixed exogenous demand, boundary conditions, or disturbances, and $f$ is a (possibly nonlinear) state-transition model. The goal is to find the sequence of inputs $\mathbf{u} = \{u_t\}_{t=0}^{T-1}$ and states $\mathbf{x} = \{x_t\}_{t=0}^{T}$ that minimize $J$, subject to the dynamics and initial conditions. We constrain $\mathbf{u} \in \mathcal{I}$, where $\mathcal{I}$ is the set of admissible control trajectories in which each control input $u_t$ acts only on vehicle speeds. This includes controlling speed limits or vehicle accelerations and excludes control inputs that vary capacity or demand, such as lane expansion. The best achievable performance under $\mathcal{I}$ is
\begin{align}
J^* \;=\; \min_{\mathbf{x},\,\mathbf{u}} \quad & J(\mathbf{x}, \mathbf{u}) \\
\text{s.t.} \quad & x_{t+1} = f(x_t, u_t, d_t) \quad \forall t \in \{0,\dots,T-1\}, \\
& x_0 = x_0^{\mathrm{obs}}, \\
& \mathbf{u} \in \mathcal{I}.
\end{align}

Let $J_{\mathrm{hist}}$ denote the value of the same metric under historical,
uncontrolled conditions. We define controllability as the best relative improvement
\begin{equation}
C \;=\; \frac{J_{\mathrm{hist}} - J^*}{J_{\mathrm{hist}}}.
\end{equation}

In this work, we focus on the setting of \textit{controllable congestion}, where $J$ represents the total delay of all vehicles in the system. Here, $J_{\text{hist}}$ is the total system delay under the simulated baseline scenario without control, while $J(\mathbf{x}, \mathbf{u})$ is the counterfactual system delay under traffic states $\mathbf{x}$ and control trajectory $\mathbf{u}$. The controllable congestion of this system is then $C$, which is an upper bound on the potential delay improvement. 

\section{Methodology} \label{met}

To evaluate controllable congestion, we want $\mathcal{I}$ to include all possible speed-based control trajectories. However, this is a large set and makes $C$ computationally difficult to solve directly. Therefore, we introduce the following assumptions in our methodology to make the computation of $C$ tractable:
\begin{enumerate}
    \item The highway has discrete segments that the control is applied to at each time step.
    \item The control input, $u_t$, is the enforced speed limit of each segment at time step $t$. We assume that dense, frequent speed limits approximate all speed-based control.
    \item The state of the system is composed of the macroscopic traffic variables at each segment.
\end{enumerate}


\paragraph{State Space} 
We consider a freeway discretized into $m$ control segments of
length $L$ each. The time domain is also discretized into time steps of length $\delta$. Let $\mathcal{M}$ denote the set of segments. Each segment
$\ell \in \mathcal{M}$ has $\lambda_\ell$ lanes and is represented by a macroscopic
traffic state
\begin{align}
    x_t(\ell) = \big[\rho_t(\ell), v_t(\ell)\big],
\end{align}
where $\rho_t(\ell)$ is the density of segment $\ell$ at time step $t$ and $v_t(\ell)$ is average velocity of vehicles in segment $\ell$ at time step $t$. The full system state at time $t$ is represented as $x_t = (x_t(1), \ldots, ~x_t(m), ~w_t)$, where $w_t$ is the queue at the mainline origin. Incorporating a queuing variable allows for analysis of open systems because it captures any effects the control may have upstream of the highway of interest. We assume all lanes have the same traffic state. 

\paragraph{Control Input}
The control input is the enforced space-mean speed limit $u_t(\ell)$ applied to all lanes of segment $\ell$ at time $t$. While speed limits do not cover all possible speed-based management strategies (such as acceleration control of AVs), densely-located and frequent speed limits provide an appropriate estimation of full control over a system. 

\paragraph{Dynamics}
The traffic state, then, evolves according to a macroscopic traffic model, $f$, such as CTM or METANET, where $x_{t+1}= f~\!\big(x_t,\, u_t,\, d_t\big)$. The exogenous inputs at time $t$ are denoted as $d_t$. This includes the mainline demand in \eqref{inflow}, the downstream boundary density, the on-ramp inflows $r(\ell)$, and off-ramp splitting ratios $\beta(\ell)$. When applied to a real corridor, the parameters of the traffic model $f$ are calibrated to observed velocities and densities (from loop detectors or vehicle trajectories) by minimizing a normalized weighted least-squares error. The full procedure is described in Section~\ref{sec:i24}.

\paragraph{Performance Metric (Total Delay)}
Congestion is measured via \emph{total delay}, defined as the excess travel time
compared to free-flow conditions. We define \emph{travel time} as the sum of both driving time and waiting time (in queue). Let $T_{FF}$ denote the total time spent (TTS) of all vehicles if they had been driving at free-flow speed, $v_{\text{free}}$. For uncontrolled conditions, the TTS is defined as $T_{\text{UC}}$, yielding uncontrolled delay $J_{\text{hist}} = T_{\text{UC}} - T_{\text{FF}}$. The quantity $J_{\text{hist}}$ can be calculated using densities from simulating the baseline calibrated model without control. Under states $\mathbf{x}$ and control trajectory $\mathbf{u}$, the controlled total travel time is
\begin{align}
T_{C}(\mathbf{x}, \mathbf{u}) = \delta \sum_{t=0}^{T} \Big(w_t \;+\; \sum_{\ell \in \mathcal{M}}
\rho_t(\ell)\,L\,\lambda_\ell\Big), 
\end{align}
and the corresponding controlled delay is $J(\mathbf{x}, \mathbf{u}) = T_C(\mathbf{x}, \mathbf{u}) - T_{FF}$. Let $\mathcal{I}_{L,\delta} \subseteq \mathcal{I}$ denote the set of speed limit trajectories on segments of length $L$ and time steps of length $\delta$. The minimum achievable delay under speed limit control at this resolution is
\begin{align} 
J^*_{L,\delta} \;=\; \min_{\mathbf{x},\,\mathbf{u}} \quad
  & T_C(\mathbf{x}, \mathbf{u}) - T_{FF} \label{eq:opt-cc}\\
\text{s.t.} \quad
  & x_{t+1} = f(x_t, u_t, d_t)
    \quad \forall \; t \in \{0,\dots,T-1\}, \\
  & x_0 = x_0^{\mathrm{obs}}, \\
  & \mathbf{u} \in \mathcal{I}_{L,\delta} \label{eq:opt-cc-end}.
\end{align}
The \emph{relaxed controllable congestion} is the corresponding relative reduction in delay,
\begin{equation} \label{eq:opt-cc-val}
C_{L,\delta} \;=\; \frac{J_{\mathrm{hist}} - J^*_{L,\delta}}{J_{\mathrm{hist}}}.
\end{equation}
Because the uncontrolled trajectory is feasible for \eqref{eq:opt-cc}--\eqref{eq:opt-cc-end} and delay is nonnegative, we have $0 \leq J^*_{L,\delta} \leq J_{\mathrm{hist}}$. It follows that $0 \leq C_{L, \delta} \leq 1$.

\noindent\paragraph{\textbf{Definition (Controllable Congestion)}} Let $L$ be the length of spatial segments (spatial resolution) and $\delta$ the length of a control time step (temporal resolution). Then, under the assumption that dense, frequent speed limits approximate all speed-based control, controllable congestion is
\begin{equation}
C \;=\; \lim_{L \to 0,\; \delta \to 0} C_{L,\delta},
\end{equation}
i.e., the maximal achievable reduction in delay as spatial and temporal resolution become arbitrarily fine. For tractability, we compute $C_{L,\delta}$ with $L$ and $\delta$ as small as the simulation allows.

\subsection{Smooth Reformulation of the System Dynamics} \label{sec:dynamics}

METANET, summarized on lines \eqref{density}--\eqref{queuing}, provides the dynamics $f$ in \eqref{metanet}. However, two of its update equations contain a min operator, which is non-differentiable and prevents the direct use of a gradient-based solver: the cap of the desired speed at the enforced speed limit in \eqref{V} and the three-term origin flow in \eqref{inflow}. There are several existing ways to handle these operators. Introducing binary variables turns the problem into a mixed-integer nonlinear program (MINLP), which can be time-intensive at long horizon lengths. Smooth approximations avoid this: for example, \citet{chavoshi} use a log-sum-exp form and report that it is faster when compared to the MINLP approach. In our setting, however, the log-sum-exp approach could result in numerical underflow at large values. We instead use the smooth minimum operator of \citet{chen1996class}, which remains numerically stable for quantities as large as traffic flows, with an explicit bound on the approximation error. For $a, b \in \mathbb{R}$ and a smoothing parameter $\varepsilon > 0$, define
\begin{align}
\mathrm{sm}^-_{\varepsilon}(a,b) := \tfrac{1}{2}\Big(a + b - \sqrt{(a-b)^2 + \varepsilon}\Big).
\label{eq:smoothops}
\end{align}

Equation \eqref{eq:smoothops} is the Chen-Harker-Kanzow-Smale smooth plus function applied to the identity $\min(a,b) = a - \max(a-b, 0)$. This smooth min operator is infinitely differentiable, since the $(a-b)^2 + \varepsilon$ radicand is bounded below by $\varepsilon > 0$, and it recovers $\min(a, b)$ as $\varepsilon \to 0$. The error is bounded uniformly and independently of $a$ and $b$,
\begin{align}
0 \;\le\; \min(a,b) - \mathrm{sm}^-_{\varepsilon}(a,b) \;\le\; \frac{\sqrt{\varepsilon}}{2}.
\label{eq:smootherr}
\end{align}

We set $\varepsilon = 10^{-4}$ throughout, so the worst-case deviation from the exact operator is $5 \times 10^{-3}$, well below any operationally meaningful model mismatch. Since $\mathrm{sm}^-_{\varepsilon}$ underestimates the minimum, the reformulation is mildly conservative. It never allows a flow or a desired speed larger than METANET would. Equation \eqref{V} accordingly becomes
\begin{align}
V[\rho_t(\ell)] =
\mathrm{sm}^-_{\varepsilon}\!\left(
v_{\text{free}}(\ell)\,
\exp\!\left[-\frac{1}{\alpha(\ell)}\left(\frac{\rho_t(\ell)}{\rho_{cr}(\ell)}\right)^{\alpha(\ell)}\right],
\;\; u_t(\ell)
\right).
\label{eq:V-smooth}
\end{align}

Equation \eqref{inflow} is smoothed by applying $\mathrm{sm}^-_{\varepsilon}$ twice in nested form:
\begin{align}
q^{O}_{t} = \mathrm{sm}^-_{\varepsilon}\!\left(
\mathrm{sm}^-_{\varepsilon}\!\left( \lambda_{1} Q \,\frac{\rho_{max} - \rho_t(1)}{\rho_{max} - \rho_{cr}(1)}, \;\; d^O_t + \frac{w_t}{\delta}\right),
\;\; \lambda_{1} Q
\right),
\label{eq:merge-smooth}
\end{align}
where the bound in \eqref{eq:smootherr} composes additively under nesting, so the three-way smoothed minimum deviates from the exact minimum by at most $\sqrt{\varepsilon}$. With these substitutions every constraint is twice continuously differentiable, the problem is a smooth --- though still nonconvex --- nonlinear problem, and the objective remains total delay as defined in \eqref{eq:opt-cc}. 

Speed limits are continuous, with $u_t(\ell) \in  [\mathcal{U}_{\min}, 150]$. The upper bound exceeds $v_{\text{free}}(\ell)$ in every segment, so $u_t(\ell) = 150$ is equivalent to imposing no speed limit. Unless stated otherwise we set $\mathcal{U}_{\min} = 0$, which is consistent with defining controllable congestion as an upper bound over \emph{all} speed-based strategies. The effect of an operationally realistic bound is examined in Section~\ref{sec:op-constraints}.

\subsection{Model Predictive Control} \label{sec:mpc}

Despite the reformulation of Section~\ref{sec:dynamics}, the problem remains highly nonconvex because the traffic model is nonlinear, and solving it directly over longer horizons (e.g., $720$ time steps for 2 hours) is intractable. We therefore use Nonlinear Model
Predictive Control (NLMPC), which decomposes the horizon into a sequence of smaller, overlapping subproblems. Let $N_p$ denote the prediction horizon and $N_c \le N_p$ the control horizon, both in time steps. At control step $k$, with $t_k = k N_c$, we solve \eqref{eq:opt-cc}--\eqref{eq:opt-cc-end} restricted to the window $\{t_k, \dots, t_k + N_p\}$ and initialized at the current state $x_{t_k}$:
\begin{equation}
\min_{\mathbf{x},\,\mathbf{u}} \quad
\delta \sum_{t=t_k}^{t_k+N_p} \Big( w_t \;+\; \sum_{\ell \in \mathcal{M}}
\rho_t(\ell)\, L\, \lambda_\ell \Big),
\end{equation}
subject to the smoothed dynamics, the constraint
$\mathbf{u} \in \mathcal{I}_{L,\delta}$, and any operational constraints. This is the same objective as \eqref{eq:opt-cc} but over the shorter window and with the constant $T_{FF}$ omitted because it does not change the minimization. The first $N_c$ actions are applied, the state is advanced by simulating the exact METANET model forward under them, and the window shifts by $N_c$; the final window is truncated to the remaining horizon. The optimizer and the simulated system share the same calibrated parameters, so there is no parametric model mismatch. Each subproblem is also given the true upstream demand and downstream density over its prediction horizon rather than a forecast of them. Therefore, they differ only through the smoothing of Section \ref{sec:dynamics}.

Control is applied on a contiguous \emph{control zone} $\mathcal{M}_c \subseteq \mathcal{M}$. For $\ell \notin \mathcal{M}_c$ we fix $u_t(\ell) = v_{\text{free}}(\ell)$ and evaluate the fundamental diagram without the speed limit in \eqref{eq:V-smooth}, so those segments evolve freely. This ensures that any delay displaced outside of the control zone is counted rather than hidden. 

Each subproblem is warm started, and the quality of that initialization largely determines whether the solver converges. A candidate speed limit trajectory is first simulated forward through METANET, so that consistent initial values are available for every variable rather than for the controls alone. We use an ordered cascade of initializations, advancing to the next only when the solver fails to return a (locally) optimal point: the previous subproblem's solution, shifted forward by $N_c$ steps, then spatially and temporally constant speed limits, beginning with the mean of the currently measured segment velocities and followed by $150$, $130$, $100$, $70$ and $50$ km/hr, and finally a cold start.

The subproblems are formulated in Pyomo and solved with IPOPT 3.14.16, a primal-dual interior-point solver, limited to $40{,}000$ iterations and $180$~s of CPU time each (although no subproblem reached either limit). The computation of each subproblem is deterministic because IPOPT returns the same solution given the same warm start and solver settings. Since the simulation is also deterministic, each scenario is run once. Unless stated otherwise we use $N_p = 45$ and $N_c = 5$, corresponding to $500$~s prediction and $50$~s control windows at $\delta = 10$~s. In the hold-length analysis of Section~\ref{sec:op-constraints}, the horizons are tied to the hold length $H$ as $N_p = 2H$ and $N_c = H$, so that each subproblem spans two complete control intervals. That sweep therefore varies the prediction horizon together with $H$. Two approximations separate the reported values from the exact optimum $C_{L, \delta}$ of \eqref{eq:opt-cc-val}: the receding-horizon truncation and the local convergence of the solver on a nonconvex program. Both act in the same direction, since any trajectory the solver returns is feasible for the dynamics and therefore achievable. The reported controllable congestion is thus a \emph{lower bound} on the $C_{L,\delta}$ \emph{upper bound} and should be interpreted as a conservative estimate of the maximum delay that speed control can remove. 

\section{Results} \label{results}

The primary objective of this section is to show how quantifying controllable congestion allows for more informed decision-making and provides information previously unknown from existing metrics. As described in Section \ref{met}, the value of controllable congestion is obtained by formulating and solving the nonlinear optimization problem defined in Equations \eqref{eq:opt-cc}–\eqref{eq:opt-cc-end}. In the following, we present results that characterize controllable congestion and evaluate its implications for traffic management. Specifically, we address two guiding questions:
\begin{enumerate}
    \item What additional insight does the controllable congestion metric provide beyond traditional metrics such as total delay?
    \item How can controllable congestion inform more targeted and effective decision-making for intelligent transportation system (ITS) infrastructure, such as VSL?
\end{enumerate}

\subsection{Case Study I: Synthetic Bottleneck} \label{sec:synthetic}

We consider a representative bottleneck scenario of length 10 kilometers, where the first 8 km are 4 lanes and the last 2 km are 2 lanes. Therefore, the roadway capacity after 8 km decreases, creating congestion when demand is high. The entire highway is split into 25 segments of 400 m (i.e. $L = 400$ meters, $m = 25$). The last 6 km, i.e. 15 segments, is the control zone where speed limit control is active. The speed limits for the first 4 km are set to the free flow speed of 120 km/hr. 

\begin{figure}[ht]
    \centering
\includegraphics[width=\linewidth]{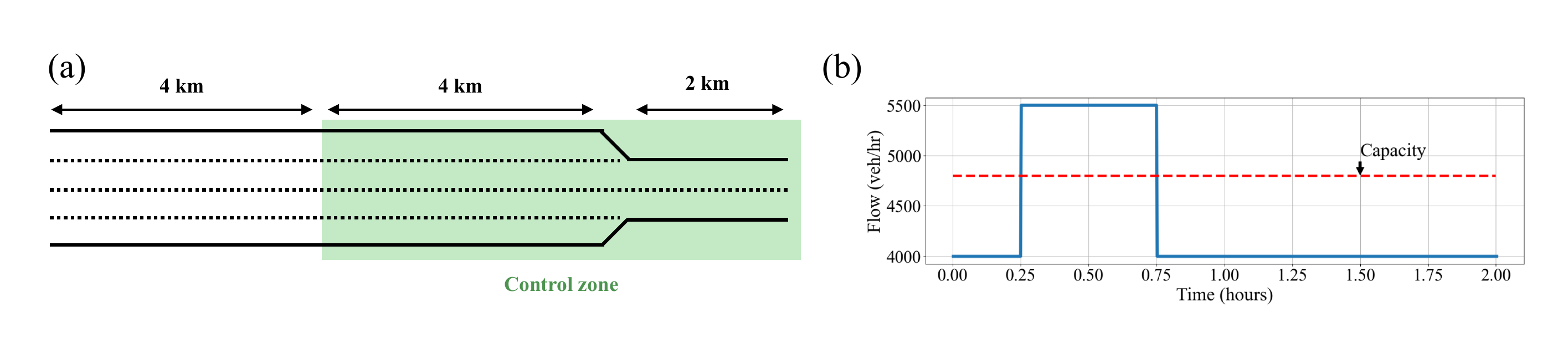}
    \caption{(a) The highway structure for the synthetic case study.
    (b) Example of demand profile at the highway entrance (in veh/hr).}
\label{fig:scenario-info}
\end{figure}

For the demand, we use a canonical scenario from \cite{chavoshi2023feedback}, where the demand increases for 30 minutes, out of the total simulation time of 2 hours, as shown in Figure \ref{fig:scenario-info}. During the increased demand, a jam is triggered. The simulation time step is set to $\delta = 10$ seconds, so $T = 720$. The parameters $\{v_{\text{free}}, \rho_{cr}, \alpha, \tau, \nu, \kappa, Q, \rho_{max}\}$ are set to default values from previous studies \citep{chavoshi2023feedback}. At these demand levels, no queue forms at the corridor entrance and all reported delay is accrued on the mainline.

\begin{figure*}[!t]
    \centering
    \includegraphics[width=\linewidth]{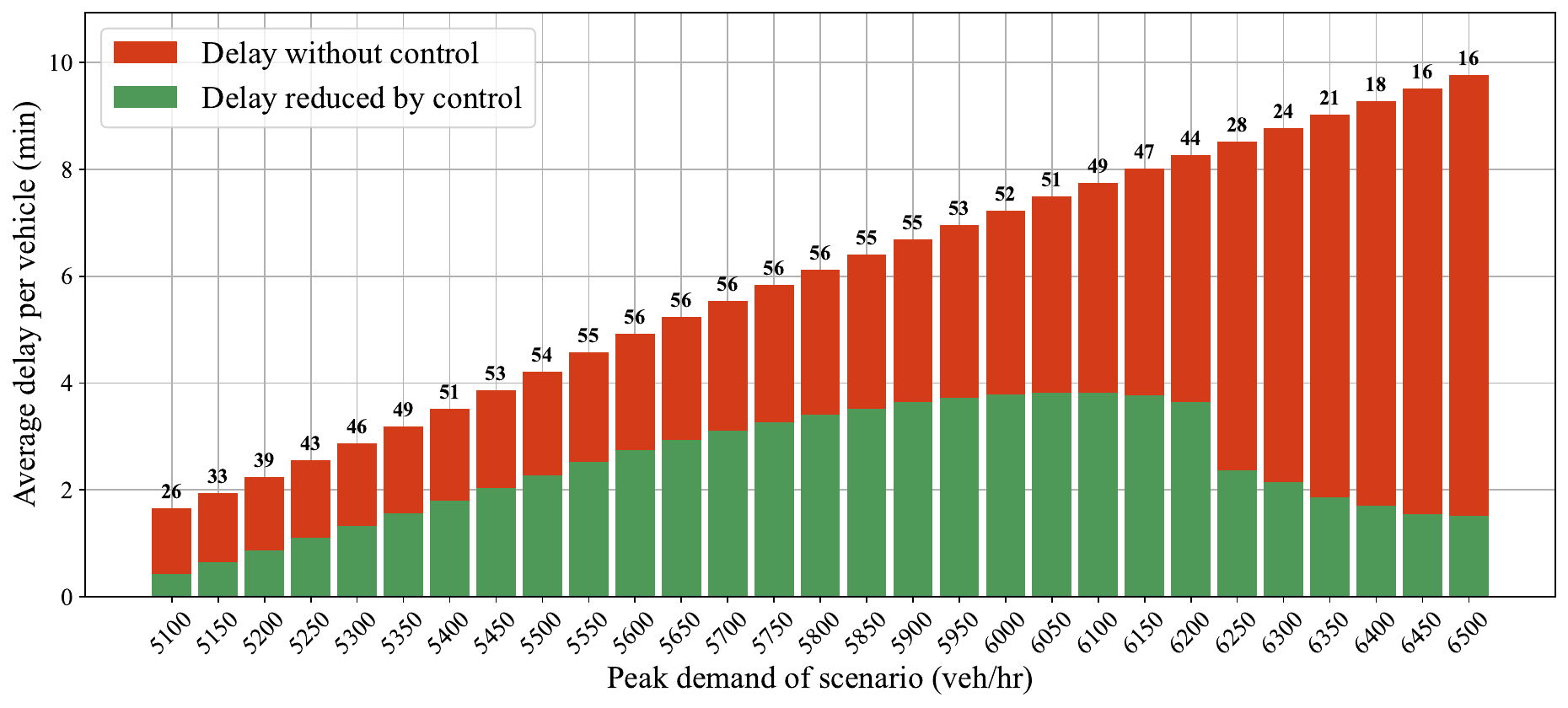}
    \caption{Average delay per vehicle for each demand scenario. The full bar
    is the delay without control, while the green portion is the delay removed by
    optimal speed limit control. Controllable congestion (\%) is printed above each bar.}
\label{fig:all-scenario}
\end{figure*}

Different scenarios are created by varying the demand during the rectangular peak within the range $[5100, 6500]$ veh/hr, in increments of 50 veh/hr. As the peak demand increases, the length and severity of the jam increases. Figure \ref{fig:all-scenario} shows, for each scenario, the average delay per vehicle without control alongside the portion of that delay which speed limit control is able to remove. While delay increases monotonically, controllable congestion (\%) rapidly increases as demand increases, plateaus at approximately 56\% between 5600 and 5800 veh/hr, and slowly decreases again at larger inflows until it decreases drastically beyond 6200 veh/hr. The results indicate that controllable congestion captures system behavior that total delay alone does not. Two corridors with 6200 and 6250 veh/hr are nearly indistinguishable by delay, level of service, or any point measure, yet the delay at the first is roughly 1.6 times as responsive to speed control as the delay at the second. Beyond this threshold, the effectiveness of speed-based control declines sharply, implying that investment in ATM strategies may not be helpful and alternative measures, such as adding capacity, should be considered. Agencies relying exclusively on delay or level-of-service metrics would likely miss this.

\begin{figure*}[!t]
    \centering
    \includegraphics[width=\linewidth]{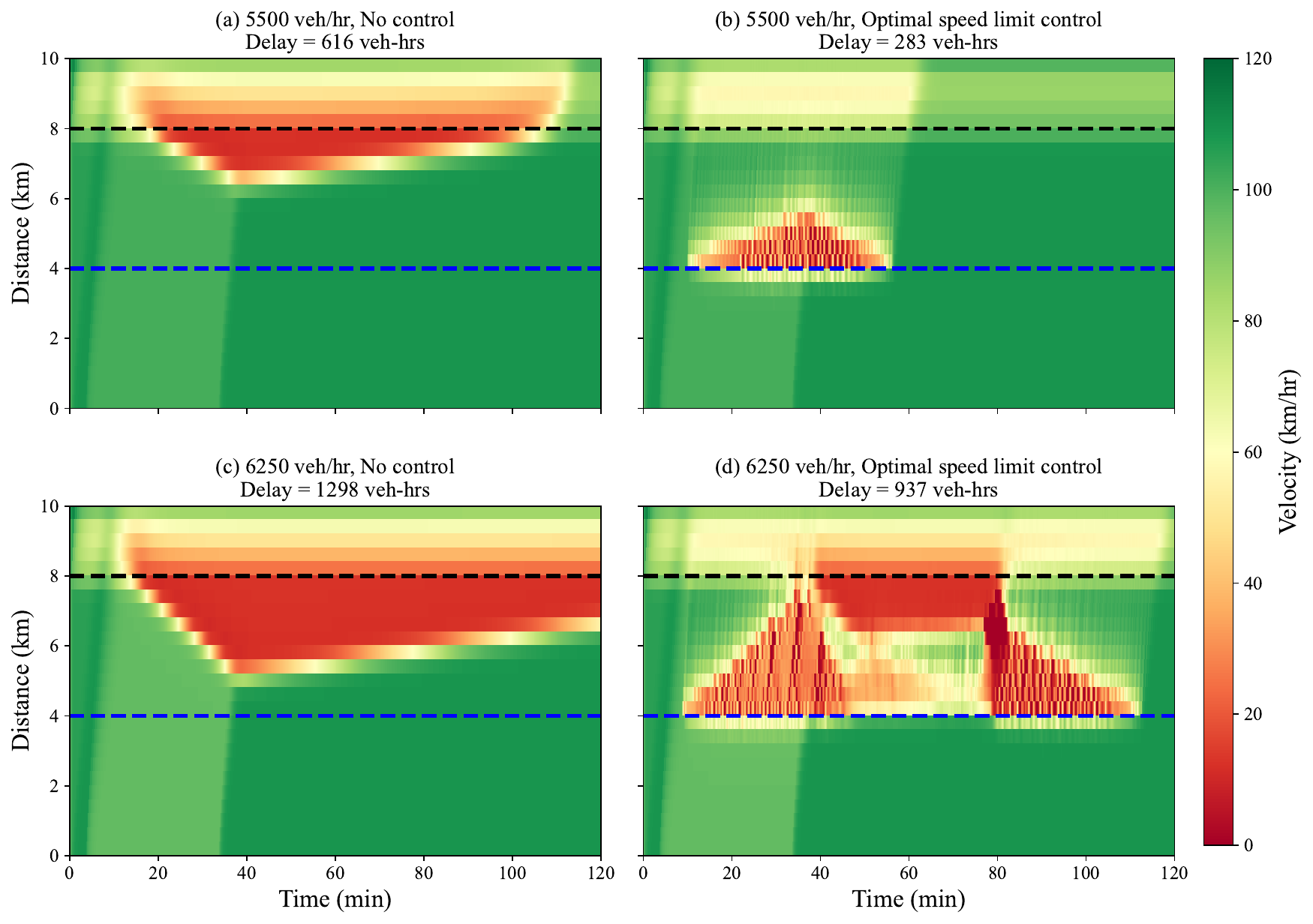}
    \caption{Velocity fields for peak demands of 5500 veh/hr (a, b) and 6250 veh/hr (c, d), without control (left) and under optimal speed limit control (right). The black dashed line indicates the bottleneck at 8 km, where the corridor narrows from four lanes to two, and the blue dashed line marks the start of the control zone at 4 km.}
\label{fig:synthetic-ts-diagrams}
\end{figure*}

The time-space diagrams --- without control and with optimized speed-limit control --- for specific peak demands of 5500 veh/hr and 6250 veh/hr are shown in Figure \ref{fig:synthetic-ts-diagrams}. For the 5500 veh/hr scenario, speed control keeps the bottleneck in free flow throughout, at the cost of a contained slowdown between 4 and 6 km (\ref{fig:synthetic-ts-diagrams}b), reducing delay from 616 to 283 veh-hrs. In particular, the optimal speed limit control mimics principles of gating control, which aims to slow vehicles down ahead of the bottleneck to prevent the capacity drop phenomenon. At 6250 veh/hr, the same strategy is applied more aggressively, yet a breakdown still occurs downstream of the bottleneck (Figure \ref{fig:synthetic-ts-diagrams}d). Control postpones it by roughly twenty minutes rather than preventing it, and delay falls only from 1298 to 937 veh-hrs.

\subsection{Case Study II: I-24 SMART Corridor} \label{sec:i24}

We now apply the same procedure to a 5.6 km stretch of the I-24 SMART Corridor near Nashville, Tennessee. Density and flow fields are constructed at a resolution of 400 m and 10 s from the available I-24 MOTION trajectory data \citep{gloudemans202324}. This results in 14 mainline segments, where the two outermost cells supply the upstream flow and downstream density boundary conditions. Each day is analyzed over a one-hour scenario during morning rush hour beginning at 08:00 am, so $T = 360$. The two segments adjacent to the inflow boundary are left uncontrolled, giving a control zone of 12 segments, or 4.8 km. We consider ten consecutive weekdays, November 21st through December 2nd 2022, two of which (November 24th and 25th) fall on the Thanksgiving holiday.

METANET parameters are calibrated separately for each day using the open-source procedure of \citet{chan2026open}. Per-segment values of $\{\tau, \nu, \kappa, \rho_{cr}, v_{\text{free}}, \alpha\}$ are estimated by solving a single nonlinear optimization problem in which the model states are decision variables and the METANET updates enter as equality constraints, minimizing a normalized weighted least-squares error against the observed velocity and density fields. Lane counts and ramp locations are retrieved from I-24 network files, and time-invariant values for on-ramp inflows and off-ramp split ratios are estimated only at cells where a ramp is present. One parameter set is fitted per day, and capacity and jam density are held fixed at 2400 veh/hr/lane and 180 veh/km/lane.

Table \ref{tab:cc_static} reports the error of the calibrated model and the resulting controllable congestion for each day. The TTS error is the mean average percent error in total time spent, the aggregate on which controllable congestion is defined. It remains below 3\% on six of the ten days and never exceeds 12\%, indicating that the calibrated model reproduces the quantity of interest for controllable congestion. We also report the root mean squared error (RMSE) of the simulated velocity field. On the eight congested days, the velocity RMSE ranges from 12.3 to 18.8 km/hr, which is comparable to or better than the 14--28~km/hr RMSE range reported in other METANET calibration studies \citep{zhao2025bounded, zhang2024discrete}.
\begin{table}[ht]
\centering
\small
\caption{Simulation error and controllable congestion by date. TTS values and controllable congestion are rounded to the nearest integer.}
\label{tab:cc_static}
\begin{tabular}{lcccccc}
\toprule
Date & \shortstack{Vel. RMSE\\(km/hr)} & \shortstack{TTS Error\\(\%)} & \shortstack{No Control\\TTS (veh-hr)} & \shortstack{With Control\\ TTS (veh-hr)} & \shortstack{Controllable\\Congestion (\%)} & \shortstack{TT Reduced\\(min/veh)} \\
\midrule
11/21 (Mon) & 15.1 & 0.6\% & 538 & 385 & \textbf{81\%} & 1.1 \\
11/22 (Tue) & 18.8 & 9.2\% & 480 & 312 & \textbf{85\%} & 1.0 \\
11/23 (Wed) & 12.3 & 10.4\% & 354 & 269 & \textbf{89\%} & 0.6 \\
\midrule
11/24 (Holiday) & 8.3 & 6.5\% & 82 & 82 & \textbf{0\%} & 0.0 \\
11/25 (Holiday) & 5.8 & 0.3\% & 132 & 132 & \textbf{0\%} & 0.0 \\
\midrule
11/28 (Mon) & 13.0 & 2.4\% & 655 & 535 & \textbf{35\%} & 0.9 \\
11/29 (Tue) & 18.2 & 1.6\% & 590 & 405 & \textbf{61\%} & 1.3 \\
11/30 (Wed) & 17.7 & 2.3\% & 568 & 392 & \textbf{65\%} & 1.1 \\
12/01 (Thu) & 14.2 & 12.0\% & 652 & 587 & \textbf{17\%} & 0.5 \\
12/02 (Fri) & 14.0 & 1.3\% & 634 & 356 & \textbf{86\%} & 2.0 \\
\bottomrule
\end{tabular}
\end{table}

Across the eight congested days, controllable congestion ranges from 17\% to 89\%, corresponding to between 0.5 and 2.0 minutes of travel time saved per vehicle for the 5.6 km I-24 SMART Corridor. The corridor accumulated 2,107 veh-hrs of total delay, of which speed limit control could have removed 1,234 veh-hrs, giving a delay-weighted controllable congestion of 58.5\%. On the two holidays, the corridor is uncongested, with delay below 0.1 minutes per vehicle, and controllable congestion is zero: there is no delay available to remove. This is the expected behavior of the metric and serves as a consistency check. 

\begin{figure}[ht]
    \centering
    \includegraphics[width=0.9\linewidth]{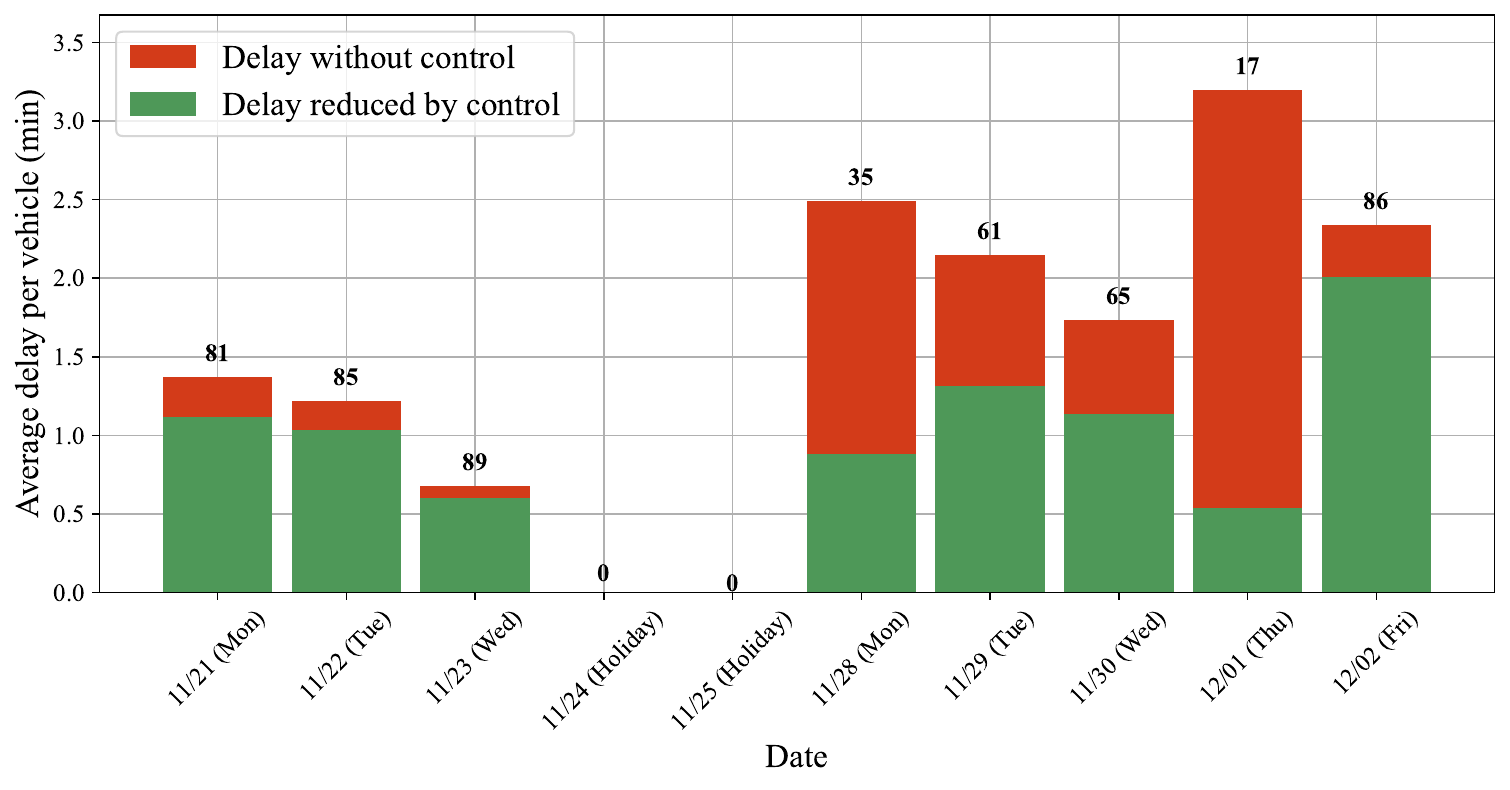}
    \caption{Average delay per vehicle for each day of the study period. The full bar is the delay without control; the green portion is the delay removed by optimal speed limit control, so the green fraction of each bar equals that day's controllable congestion, printed above the bar.}
\label{fig:i24-bar}
\end{figure}

\begin{figure}[ht]
    \centering
    \includegraphics[width=\linewidth]{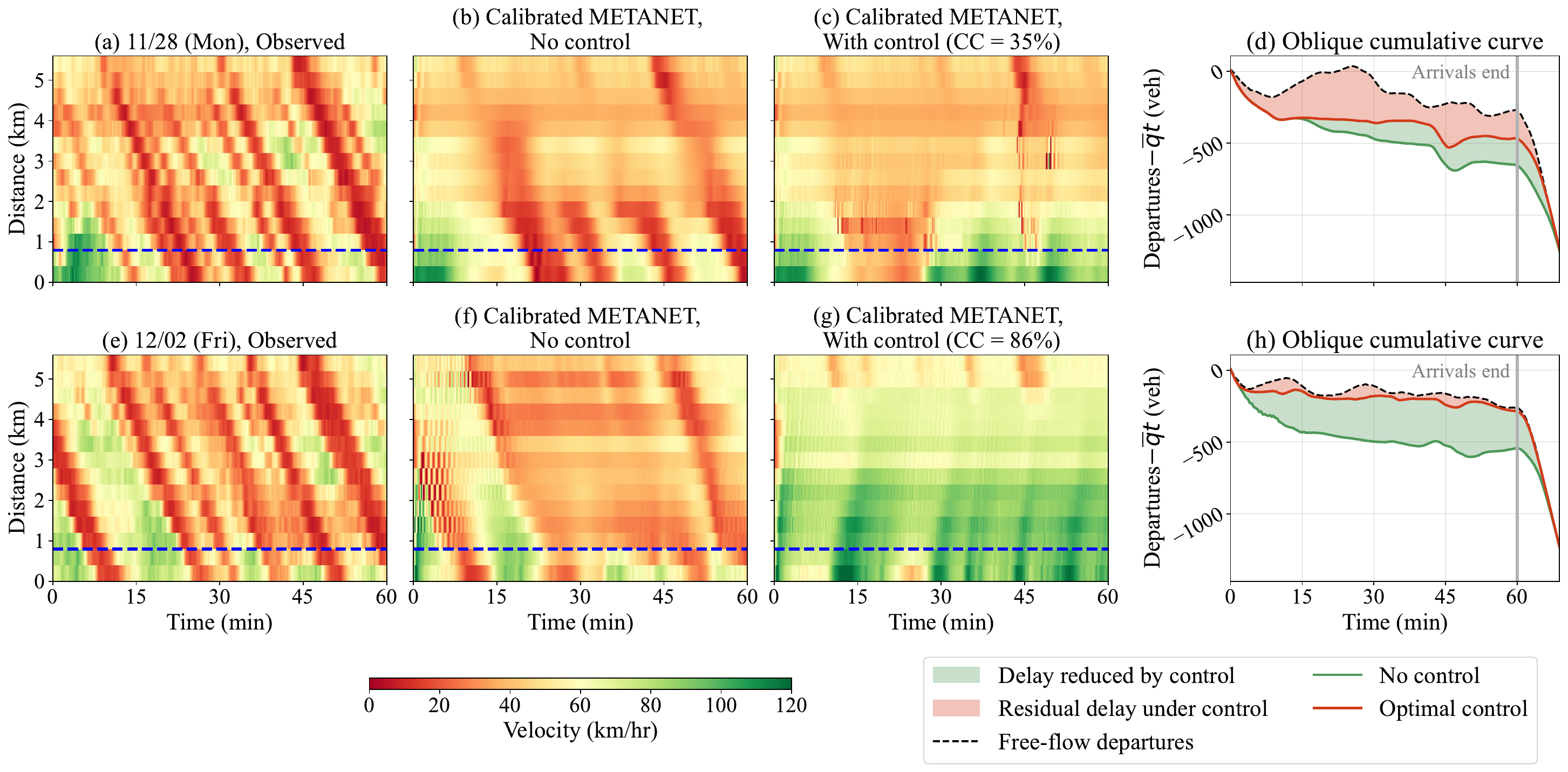}
    \caption{Velocity fields for Monday, November 28th (a -- c) and Friday, December 2nd (e -- g): observed (left), without control (middle) and with speed limit control (right). The blue dashed line marks the start of the control zone at 0.8 km. Rightmost panels (d) and (h) show cumulative departures minus $\overline{q} t$, where $\overline{q}$ is the average arrival rate of vehicles per hour (8,373 and 8,326 veh/hr for Mon and Fri, respectively), for free flow, with control, and no control. Delay can be seen visually as the area between the dashed line and green or red lines.}
\label{fig:i24-tsd}
\end{figure}

Figure \ref{fig:i24-bar} shows that, as in the synthetic study, \emph{delay and controllable congestion are largely unrelated}. Monday, November 28th and Friday, December 2nd are almost indistinguishable by conventional measures, with a TTS of 655 and 634 veh-hrs, respectively. Controllable congestion, however, was 35\% on the Monday and 86\% on the Friday, more than \emph{twice} as large. The time-space diagrams and cumulative departure plots for both of these days, shown in Figure \ref{fig:i24-tsd}, indicate that the difference lies not in where congestion occurs but in whether control can clear it. On November 28th, speed control improves the average speed in the upstream half of the corridor from 44 to 65 km/hr while leaving the downstream half nearly unchanged. On December 2nd, control clears both upstream and downstream congestion, evenly reducing delay across the entire hour and tracking the free-flow departures much closer. Having both delay and controllable congestion as metrics can better inform where congestion is more responsive to speed control.

\subsection{Operational Constraints on Controllable Congestion} \label{sec:op-constraints}

The controllable congestion reported in Section \ref{sec:i24} is an upper bound and is computed under the full intervention set, with speed limits free to vary at every segment and every time step. In practice, highway operators face restrictions on how frequently speed limits may change, how sharply they may differ across space and time, and how low they may be posted \citep{katz2012guidelines}. A speed control strategy that ignores these restrictions is not deployable, so a natural question is how much of the controllable congestion remains once they are imposed. In this section, we introduce four such constraints, add each to the optimization problem described in (\ref{eq:opt-cc}) -- (\ref{eq:opt-cc-end}), and measure the resulting controllable congestion as the constraint is tightened. All results are reported for November 30th, whose unconstrained controllable congestion is 65.4\%.

\paragraph{Hold constraint}
In many settings, speed limits cannot be updated at every time step, either due to driver safety or latency of a deployed system. Let the \emph{hold length} be \(H\in\mathbb{N}\), meaning the control must remain
constant for \(H\) consecutive time steps. If there is a hold length, the following constraint will need to be enforced:
\begin{align}
    \text{(Hold length)} \quad & u_{t}(\ell) = u_{t+1}(\ell),
    && \forall\, \ell\in\mathcal{M}_c,\;
    \forall\, t\in\{0,\dots,T-2\}\setminus\{H-1,~2H-1,\dots\} \label{eq:holdlen}
\end{align}

\paragraph{Minimum speed limit constraint}
It is considered unsafe to control vehicles to go too slow, especially if a human driver is behind the wheel. Often, many systems, like VSL, also have legal requirements on the minimum displayable speed that they must comply with. To account for this, we impose a lower bound on the speed limit control, denoted as  \(\mathcal{U}_{\min}\) \(\geq 0\),
\begin{equation}\label{eq:umin}
u_t(\ell)\;\ge\; \mathcal{U}_{\min},
\qquad \forall\, t\in\{0,\dots,T-1\},\; \forall\, \ell\in\mathcal{M}_c.
\end{equation}

\paragraph{Safety / smoothness constraint}
Finally, we want to encourage speed control that is smooth, both over time and space, to prevent sudden accelerations and decelerations that could cause accidents or driver panic. Let \(\mathcal{E} := \{(\ell,\ell')\in\mathcal{M}_c\times\mathcal{M}_c:\ell'=\ell+1\}\) denote adjacent segment pairs. Given spatial and temporal upper bounds
\(\mathcal{S}_{\mathrm{spat}}, \text{ } \mathcal{S}_{\mathrm{temp}}\ge 0\),
\begin{subequations}\label{eq:safety}
\begin{align}
\text{(Spatial safety)}\quad
& \bigl|u_t(\ell)-u_t(\ell')\bigr| \le \mathcal{S}_{\mathrm{spat}},
&& \forall\, t\in\{0,\dots,T-1\},\; \forall\, (\ell,\ell')\in\mathcal{E},
\label{eq:spatial}
\\
\text{(Temporal safety)}\quad
& \bigl|u_{t}(\ell)-u_{t+1}(\ell)\bigr| \le \mathcal{S}_{\mathrm{temp}},
&& \forall\, t\in\{0,\dots,T-2\},\; \forall\, \ell\in\mathcal{M}_c.
\label{eq:temporal}
\end{align}
\end{subequations}

\begin{figure*}[!t]
    \centering
    \includegraphics[width=\linewidth]{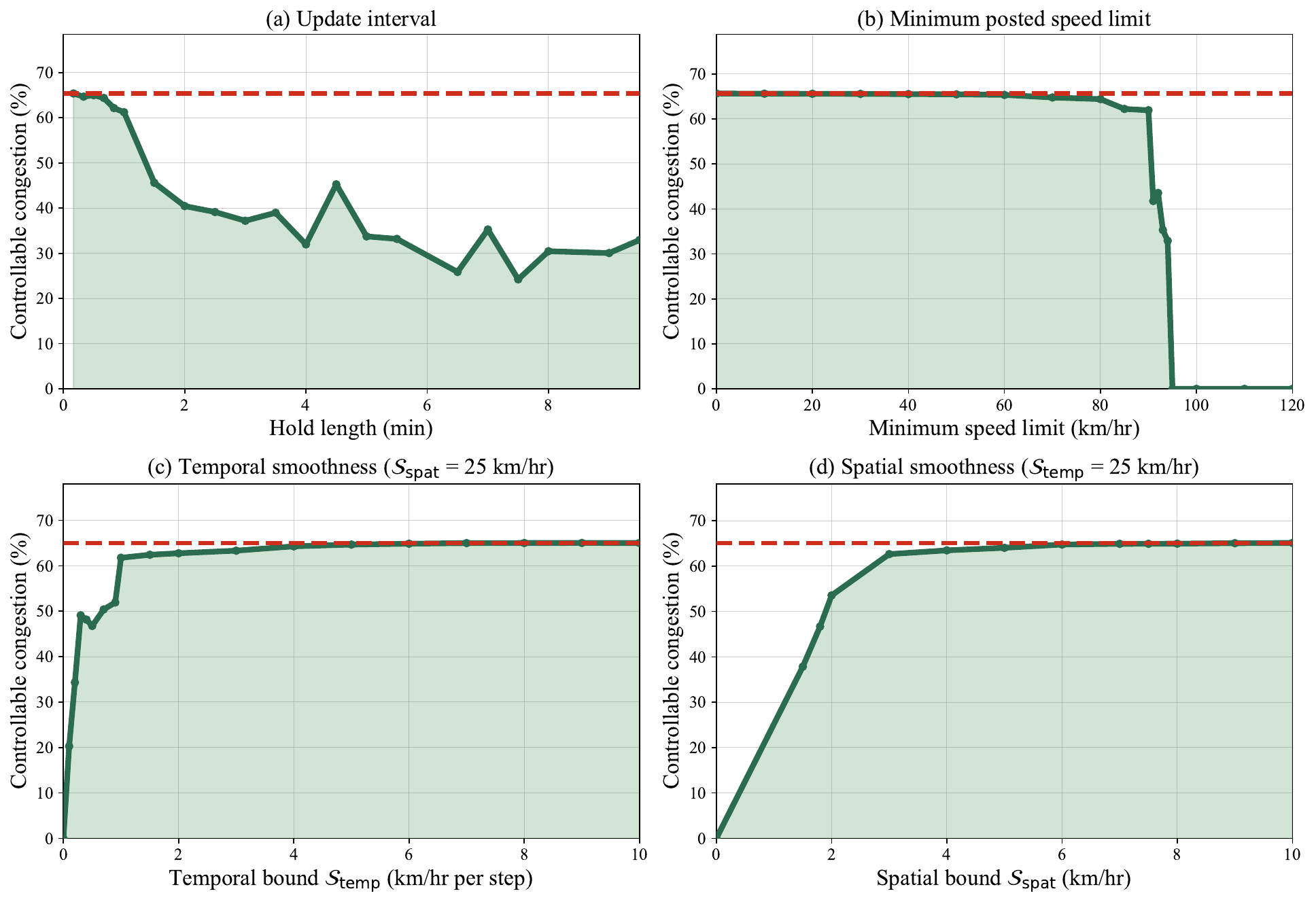}
    \caption{Controllable congestion on November 30th under operational constraints. (a) Hold length, from 10 seconds to 10 minutes. (b) Minimum posted speed limit, from 0 to 120 km/hr. (c) Temporal smoothness bound, at a fixed spatial bound of 25 km/hr. (d) Spatial smoothness bound, at a fixed temporal bound of 25 km/hr. The red dashed line in each panel marks the unconstrained controllable congestion of 65.4\%. Some of the trends are not monotonic and show more variability due to the locally optimal nature of MPC.}
    \label{fig:sensitivity}
\end{figure*}

Each constraint is added to the optimization problem solved at every MPC step and swept independently, with the remaining three left inactive. Figure \ref{fig:sensitivity} reports the resulting controllable congestion. Some of the trends shown are not monotonic and have variability due to the locally optimal nature of MPC. Panel (a) shows that controllable congestion is essentially unaffected by hold lengths up to approximately one minute, retaining a controllable congestion of 61\%. Beyond one minute it falls sharply, dropping to 40\% at a two-minute hold, and decays gradually thereafter to roughly half of the unconstrained value at hold lengths of 2 -- 10 minutes. The benefit of speed control on this corridor therefore derives largely from adjustments made on a sub-minute timescale. In this sweep, the prediction and control horizons are tied to the hold length as $N_p = 2H$ and $N_c = H$ (refer to Section \ref{sec:mpc}), so the reported trend reflects the combined effect of a coarser update interval and a correspondingly shorter optimization horizon.

Figure \ref{fig:sensitivity}b shows that controllable congestion is largely insensitive to the minimum speed limit across the range an operator would plausibly consider. Raising $\mathcal{U}_{\min}$ from 0 to 50 km/hr costs less than one percentage point, and a floor as high as 90 km/hr still retains a controllable congestion of 61.9\%. The delay reduction achievable on this corridor therefore does not depend on slowing traffic to a near-standstill. Between 90 -- 95 km/hr, however, controllable congestion declines rapidly and collapses to zero beyond 95 km/hr. This transition follows directly from the calibrated free-flow speeds. A speed limit can only reduce the equilibrium speed where $\mathcal{U}_{\min} < v_{\text{free}}(\ell)$, and within the control zone only four segments have a calibrated free-flow speed above 90 km/hr. Of these, just one lies in the upstream part of the congestion, with $v_{\text{free}}(\ell = 5) = 94.2$ km/hr. As the lower bound approaches that value, the range available for the speed limit control becomes minimal, and once the lower bound exceeds it, the only remaining actuators are downstream of the congestion, where slowing vehicles cannot relieve congestion upstream of them.

Finally, panels (c) and (d) of Figure \ref{fig:sensitivity} report controllable congestion over the grid of spatial and temporal smoothness bounds. Tightening the temporal bound from 10 to 1.5 km/hr per time step reduces the controllable congestion by 2.6 percentage points, while tightening the spatial bound over the same range reduces controllable congestion by 27.3 percentage points. This highlights that large spatial differences in speed limit control provide more benefit than large temporal differences and remains consistent with the gating mechanism identified in Section \ref{sec:synthetic}. The control acts by establishing a speed differential between the upstream stretch
and the segments downstream of it. One note is that the two bounds are expressed in different units, per 10-second step in time and per 400-meter segment in space, so a nominal value of 1 km/hr is not equally restrictive on the two
axes. For a VSL operator, the implication is
that the spatial resolution of a deployment --- gantry spacing and the speed step permitted between adjacent displays --- impacts achievable benefit more than the rate at which those displays may be updated.



\section{Conclusion} \label{conc}
Performance metrics have long been used to assess highway system performance, yet metrics used by stakeholders are often not actionable and do not capture to what extent the system can be influenced by emerging systems of traffic control. To address this, we introduce controllable congestion, a metric that quantifies the upper bound of improvement attainable from speed-based traffic management, and estimate it by solving a nonlinear optimization problem over the METANET dynamics using model predictive control. 

The presented results highlight that controllable congestion captures dynamics not reflected in conventional measures such as delay. In the synthetic bottleneck, controllable congestion plateaus near 56\% and then collapses beyond a peak demand of 6200 veh/hr, while delay increases monotonically throughout. On the I-24 Corridor, it ranges from 17\% to 89\% across eight congested days, and two mornings that are nearly indistinguishable by total travel time, total delay, and delay per vehicle differ by more than a factor of two in how much of that delay speed control can remove. We further show that the realizable share of this bound depends on the operational constraints within which control is exercised. On I-24, the minimum posted speed limit has little effect on controllable congestion, while frequency of update and spatial smoothness have large impacts. These insights highlight that controllable congestion is a \emph{complementary} metric to standard delay-based metrics and will better guide stakeholders on how to effectively choose large-scale traffic control infrastructure.

The framework is not specific to congestion. The controllability optimization problem is posed over a generic performance measure $J$, so the same construction applies to any metric that can be expressed as a function of the system trajectory. Replacing total delay with an emissions or surrogate safety measure yields controllable emissions or controllable safety, estimated by the same procedure over the same intervention set. The intervention set itself can likewise be broadened beyond speed limits to include ramp metering or vehicle-level acceleration control, which would raise the bound. We therefore view controllable congestion as the \emph{first instance of a broader class of controllability-aware performance measures}.

While these results establish the value of controllable congestion, several limitations remain. Computing this metric currently requires solving many nonconvex and nonlinear optimization problems, so the reported values are achievable but locally optimal and are best interpreted as conservative estimates of the true upper bound. Future work will include addressing this via improved warm-starting and conducting an analysis on the sensitivity of the metric to the local optima found. Additionally, the metric relies on the calibrated model, so future work on improving the calibration procedure would build further trust in the outcome. METANET also does not capture all aspects of driver behavior, such as lane changing, which may influence the effectiveness of speed limit control and limit transferability from simulation to real-world settings. Finally, the empirical analysis covers ten days on a single corridor. Applying the metric across many corridors would establish how widely controllable congestion varies and how reliably it can be estimated from broader traffic data. Addressing these limitations represents an important direction for future work, with the potential to enhance both the efficiency and the realism of the controllable congestion methodology.

\section*{Acknowledgements}
This work was funded by the NSF Graduate Research Fellowship under Grant No. 2141064 and Cintra, S.A. The authors would like to thank Jason Soria and Jennifer Duthie for the numerous insightful discussions.

\section*{Author contributions: CRediT}
\textbf{Shreyaa Raghavan}: Conceptualization, Methodology, Validation, Visualization, Formal analysis, Writing – original draft. \textbf{Edgar Ramirez-Sanchez}: Conceptualization, Methodology. \textbf{Monica Chan}: Validation. \textbf{Zhengbing He}: Supervision, Writing – review \& editing. \textbf{Cathy Wu}: Conceptualization, Supervision, Funding acquisition, Writing – review \& editing.


\bibliographystyle{model2-names}
\bibliography{main}

\end{spacing}
\end{document}